\documentclass[12pt, a4paper]{article}
\usepackage[left=1.5cm, right=1.5cm, top=2cm, bottom=2cm]{geometry}
\usepackage[square,numbers]{natbib}

\usepackage{amsmath,amssymb,amsthm}
\usepackage{graphicx}
\usepackage{hyperref}
\usepackage{algorithm,algpseudocode}
\usepackage{booktabs}
\usepackage{braket}
\usepackage{siunitx}

\title{Hypertriton states from inverse scattering theory}
\author{Emile  Meoto$^1$  \and Mantile L. Lekala$^2$ }
\date{$^1$ Department of Physics, University of Buea, P.O. Box 63 Buea, South West Region, Cameroon, \\
Email: Meoto.Emile@ubuea.cm or EMeotoson@gmail.com \\
$^2$ Department of Physics, University of South Africa, Florida Park, P.O. Box 392, South Africa, \\
Email: lekalml@unisa.ac.za \\
19 July 2026}
\begin{document}

\maketitle

\begin{abstract}
The primary goal of this paper is to demonstrate that inverse scattering theory is a viable method for the simulation of lambda-nucleon potentials in hypernuclear few-body studies. To this end, we investigate the hypertriton, modelled as a $\Lambda np$ three-body system in the $J^\pi = 1/2^+$ and $3/2^+$ channels. This three-body problem is solved using a hyperspherical-harmonic expansion of the Faddeev equations. The $\Lambda p$ and $\Lambda n$ interactions are modelled by the GLM-YN0 potentials. These simulated potentials were recovered through Gel'fand--Levitan--Marchenko inverse scattering theory as phase-equivalent simulations of the NSC97f meson-exchange model. The neutron-proton interaction is described by the semi-realistic Malfliet--Tjon I/III potential, with both singlet and triplet channels retained. For the ground state ($J^\pi = 1/2^+$), we obtain a binding energy of $-2.335$~MeV, corresponding to a $\Lambda$ separation energy of $B_\Lambda = +0.104$~MeV relative to the $\Lambda + d$ breakup threshold. This value is comparable to those from lambda-nucleon potentials that are simulated through G-matrix methods. The excited $J^\pi = 3/2^+$ state is found to have a lambda separation energy of $B_{\Lambda}=-1.444 $~MeV relative to the $\Lambda + d$ breakup threshold, confirming that there is no bound excited hypertriton state.
\end{abstract}

\section{Introduction}

In hypernuclear physics, the lambda-nucleon-nucleon ($\Lambda N N$) three-body problem plays a pivotal role. Its importance is akin to that of the two-nucleon ($NN$) and three-nucleon ($NNN$) systems in the non-strange sector of nuclear physics. The deuteron ($np$) and triton ($pnn$), complemented by a rich experimental scattering database, serve as testing grounds for the nucleon-nucleon force. In the strangeness sector, the spin-isospin channel $\Lambda n p$ ($I=0$, $J^{\pi}=1/2^+$) is traditionally used for constraining the lambda-proton ($\Lambda p$) and the lambda-neutron ($\Lambda n$) forces. The key difference here is that unlike the NN case, there is no reliable lambda-nucleon scattering database to complement hypertriton calculations. The difficulties surrounding lambda-nucleon scattering experiments have been exhaustively discussed in the literature and will not be repeated here.

In the absence of reliable scattering data, the structural properties of lambda hypertriton ($\Lambda n p$) are the only observables that can be used to constrain both the $\Lambda p$ and $\Lambda n$ forces. Experimentally measured observables include its binding energy, lifetime, pion momentum and branching ratios of its mesonic decay, etc. It is of prime importance in theoretical studies that the same set of $\Lambda p$ and $\Lambda n$ forces is used to investigate all $\Lambda n p$ states, as well as the elusive $\Lambda n n$ state. This strategy offers some clarity on the following question, which is of central importance in hypernuclear physics: what $\Lambda n p (J^{\pi}=1/2^+)$ binding energy is consistent with a virtual state or resonance (or  bound state) in $\Lambda n p (J^{\pi}=3/2^+)$ and in $\Lambda n n (J^{\pi}=1/2^+)$ \cite{oga1980, miy1995, hiy2014, sch2020, sch2022}. The pair of channels $\Lambda n p (J^{\pi}=1/2^+)$  and $\Lambda n n (J^{\pi}=1/2^+)$ have frequently been studied together. There is a subtle difference between $\Lambda np$ and $\Lambda nn$: antisymmetrisation as required by the Pauli exclusion principle forbids a nn 3s1 force, thereby restricting the $nn$ force to the 1s0 state, as opposed to the np force where both 1s0 and 3s1 np channels contribute. It is the stronger 3s1 np force that renders the hypertriton a bound state. The 1s0 nn interaction is attractive but not strong enough to bind; it only produces a virtual (anti‑bound) state just below threshold \cite{hag2004}. The absence of a bound dineutron analogue to the deuteron is a key factor that does not favour a $\Lambda nn$ bound state.

The spectroscopy of $\Lambda n p$ and $\Lambda n n$ systems is currently of interest to many experimental groups \cite{rap2023}. Most recently, the A1 Collaboration at the Mainz Microtron (MAMI) determined the $\Lambda$ separation energy of the hypertriton with unprecedented precision through decay-pion
spectroscopy ~\cite{kin2026}.  Their results reveal that $^3_\Lambda\mathrm{H}$ is significantly more strongly bound than many earlier experiments suggested. By measuring the monochromatic $\pi^-$ momentum from the two-body mesonic decay $^3_\Lambda\mathrm{H} \rightarrow
\,^3\mathrm{He} + \pi^-$, the following value was
obtained:

\begin{equation}
B_\Lambda(^3_\Lambda\mathrm{H}) = 0.523 \pm 0.013_{\rm stat.} \pm 0.075_{\rm syst.}
\ \mathrm{MeV}
\label{eq:BL_MAMI}
\end{equation}

This result is consistent with the STAR measurement of $B_\Lambda = 0.406 \pm
0.120_{\rm stat.} \pm 0.110_{\rm syst.}$~MeV~\cite{ada2020a}, but indicates significantly deeper binding than the J-PARC E07 emulsion value of $B_\Lambda =
0.23 \pm 0.11_{\rm stat.} \pm 0.05_{\rm syst.}$~MeV~\cite{kas2025} and the ALICE value of $B_\Lambda = 0.102 \pm 0.063_{\rm stat.} \pm 0.067_{\rm syst.}$~MeV~\cite{ach2023}. The J-PARC E07 and ALICE values sit at the shallow end, while the MAMI A1 and STAR values are at the deeper end.

Over the last seven decades, a number of calculations on $\Lambda n n$ predicted that a bound state is not possible \cite{dal1958a, dal1958b, dow1959, gar1987,bel2008}. After a signal for a possible bound state was inferred from heavy-ion experiments by the HypHI Collaboration at GSI \cite{rap2013}, a number of recent theoretical studies, based on improved $\Lambda N$ forces from Meson-Exchange theory, Chiral Effective Field theory and quantum chromodynamics, further argued against the existence of a $\Lambda n n$  bound state \cite{gal2014, gar2014, hiy2014, afn2015, kam2016, fil2016, gib2019}. The deeper binding energy reported by the A1 Collaboration~\cite{kin2026} shifts
the theoretical landscape in a direction more favourable to a near-threshold three-body virtual state or three-body resonance in
$\Lambda nn$ \cite{afn2015}. 

Due to the complex nature of bare meson exchange potentials some few-body problems may be computationally very expensive or intractable. These bare potentials cannot be applied directly inside a nucleus because of their strong short-range repulsion. The G-matrix method is used to construct an effective interaction that addresses the hard core repulsion and also incorporates in-medium nuclear effects. For several decades, G-matrix methods have been used extensively to simulate meson-theory lambda-nucleon potentials for few-body calculations \cite{lan1997, rij1999, yam2010}. An alternative perspective on the $\Lambda N$ force was proposed by applying inverse scattering theory on NSC97f scattering phases to recover GLM-YN0 potentials \cite{meo2019, meo2020, meo2024}. Whereas G-matrix potentials depend on the density of the nuclear medium (or the Fermi momentum), inverse scattering theory is fundamentally different in its approach: it constructs medium-independent potentials that are based entirely on free-space scattering information. Another strength of GLM-YN0 potentials is that they inherit the full on-shell behaviour of the original NSC97f potentials, since they are phase-equivalent by construction. 

Whereas G-matrix potentials have the limitation that the Fermi momentum may not be accurately known, inverse scattering theory potentials suffer from the fact that off-shell behaviour is not completely encoded. However, as demonstrated in \cite{ger1990}, off-shell behaviour is reproduced in inverse scattering potentials, with variations only expected very far off the energy shell. The application of potentials constructed through inverse scattering theory has a history in the non-strange sector of nuclear physics \cite{ada1993, gib1995}. 

The aim of this paper is to study the $\Lambda n p$ ($J^{\pi}=1/2^+$) and $\Lambda n p$ ($J^{\pi}=3/2^+$) three-body problems using GLM-YN0 $\Lambda p$ and $\Lambda n$ potentials. This study aims to assess the extent to which the off-shell behaviour implicitly encoded in inverse-scattering-theory potentials is sufficient to reproduce few-body binding energies, which are known to be sensitive to the off-shell properties of the underlying two-body interaction. The computations are performed using the coupled hyperradial equations obtained from the hyperspherical harmonic expansion of Faddeev equations.

\section{Hyperspherical harmonics expansion on Faddeev equations} 

For a three-body problem, after separation of the centre-of-mass motion, the internal dynamics is described by two mass-scaled Jacobi vectors that are given in spectator notation by $(\vec{x}_i, \vec{y}_i)$. From this six-dimensional Jacobi coordinate system, six-dimensional Delves hyperspherical coordinates ($\rho$, $\theta_i$, $\nu_{x_i}$,  $\omega_{x_i}$, $\nu_{y_i}$, $\omega_{y_i}$) are constructed. The coordinate \(\rho \in [0, \infty)\) is the hyperradius, a collective coordinate representing the overall size of the three-body system. It is defined as \(\rho = \sqrt{x_i^2 + y_i^2}\) and is invariant under transformations between Jacobi coordinate sets. The coordinate $\theta_i \in [0, \pi/2]$ is the hyperangle and is defined through the parametrisation $x_i= \rho \sin \theta_i$ and $y_i = \rho \cos \theta_i$.

The pair of coordinates $\Omega_{x_i}=(\nu_{x_i},  \omega_{x_i})$ are spherical polar angles associated with the Jacobi vector $\vec{x}_i$ while $\Omega_{y_i}=(\nu_{y_i}, \omega_{y_i})$ are spherical polar angles associated with $\vec{\lambda}_i$. 

Two-body wavefunctions in Jacobi coordinates $\psi_i(\vec{x}_i, \vec{y}_i)$ are written as $\psi_i(\rho, \Omega^i_5)$ in hyperspherical coordinates. This wavefunction is expanded on a basis of hyperspherical functions as follows:

\begin{align}
\label{eq:expansion}
    \psi_{i}^{J\alpha_{i}}(\rho,\Omega_{5}^{i})=\sum_{K_{i}}\frac{\chi_{K_{i}}^{i,J\alpha_{i}}(\rho)}{\rho^{5/2}}\mathcal{Y}_{K_{i}}^{\ell_{x_{i}}\ell_{y_{i}}}(\Omega_{5}^{i})
\end{align}

where \(\chi^{i,J\alpha_i}_{K_i}(\rho)\) is the hyperradial wavefunction for channel \(i\), \(J\) is the total angular momentum of the three-body system, \(\alpha_i\)  represents the channel quantum numbers (e.g., spin, orbital angular momentum coupling), and \(K_i\) is a hyperangular momentum. The superscript \(i\) indicates the spectator particle (Jacobi set \(i\)). $\mathcal{Y}_{K_{i}}^{\ell_{x_{i}}\ell_{y_{i}}}(\Omega_{5}^{i})$ are the hyperspherical harmonics for the channel $(i, K_i, \ell_{x_{i}}, \ell_{y_{i}})$, with orbital angular momenta $\ell_{x_{i}}$ and $\ell_{y_{i}}$. $\Omega_{5}^{i}$ represents the five angles ($\theta_i$, $\nu_{x_i}$,  $\omega_{x_i}$, $\nu_{y_i}$, $\omega_{y_i}$) for the $i$-th Jacobi set.  

The wavefunction expansion in Eq. \eqref{eq:expansion} is substituted into the Faddeev equations that are expressed in Jacobi coordinates. This equation is then projected onto the complete hyperspherical harmonics basis, yielding a system of coupled hyperradial equations \cite{meo2026}

\begin{equation}
\left[-\frac{\hbar^{2}}{2m}\frac{d^{2}}{d\rho^{2}}+L_{K_{i}}-E\right] \chi^{i,J\alpha_{i}}_{K_{i}}(\rho)+\sum_{n,K_{n}}V^{in}_{K_{i}K_{n}}(\rho)\chi^{n,J\alpha_{n}}_{K_{n}}(\rho)=0 ,
\end{equation}

where \(\hbar\) is the reduced Planck constant, \(m\) is a reference mass introduced to scale the Jacobi coordinates (in nuclear physics, this is often taken as the neutron mass \(939.6\,\text{MeV}/c^2\)), \(E\) is the total energy of the three-body system (internal energy, since the centre-of-mass motion has been separated out) and \(n\) is an index labelling different Faddeev components corresponding to different Jacobi coordinate sets (\(n = i, j, k\)). \(L_{K_i}\) is a centrifugal barrier, while \(V^{in}_{K_i K_n}(\rho)\) is the coupling potential between hyperradial channels \((i, K_i)\) and \((n, K_n)\). In order to solve the coupled hyperradial equation, the hyperradial wavefunctions are expanded on a basis as follows \cite{tho2004}

\begin{align}
    \chi^{i,J}_{\alpha_{i}K_{i}}(\rho)=\sum_{n=0}^{N_{b}}a^{in,J}_{K_{i}\alpha_{i}}R_{n}(\rho)\;.
\end{align}
The basis functions are constructed as

\begin{align}
    R_{n}(\rho)=\rho^{5/2}\rho_{0}^{-3}[n!/(n+5)!]^{1/2}L_{n}^{5}(z)\exp(-z/2)\;,
\end{align}
where $z=\rho/\rho_0$, $\rho_0$ being a scaling hyperradius, and $L_{n}^{5}(z)$ is an associated Laguerre polynomial.

\section{States of the $\Lambda NN$}

The three particles $\Lambda$, $n$ and $p$ each has a spin of 1/2. The states of $\Lambda np$ are constructed by first coupling any two particles, and then using the result to couple to the third. For the $L=0$ case of interest here, it follows $J=L+S=S$. Therefore the states of the three particles are $j_{\Lambda}=s_{\Lambda} $, $j_n=s_n$ and $j_p=s_p$. From angular momentum algebra, two particles with states $j_1$ and $j_2$ combine to give the possible states $|j_1-j_2|, |j_1-j_2|+1, ..., j_1 +j_2$. Therefore, combining any two from the three particles $\Lambda$, $n$ and $p$ gives rise to

\begin{align}
    \frac{1}{2}\otimes\frac{1}{2}=0\oplus 1
\end{align}

That is, the total spin is either 0 or 1. Using the result to couple to the third particle, one arrives at

\begin{align}
   (0\oplus 1)\otimes\frac{1}{2}
=
(0\otimes\frac{1}{2})\oplus(1\otimes\frac{1}{2}) 
\end{align}

\begin{align}
    0\otimes\frac{1}{2}=\frac{1}{2},
\qquad
1\otimes\frac{1}{2}=\frac{1}{2}\oplus\frac{3}{2}
\end{align}

Therefore,
\begin{align}
    (0\oplus 1)\otimes\frac{1}{2}
=
\frac{1}{2}\oplus\left(\frac{1}{2}\oplus\frac{3}{2}\right)
\end{align}

It may be seen that the three particles can be in two independent $J=1/2$ states or one $J=3/2$ state. These are the possible states of the $\Lambda np$ and $\Lambda nn$. Each of the $J=1/2$ states is a doublet ($\ket{JM_J}$  = $\ket{1/2,-1/2}$, $\ket{1/2,1/2}$) while the $J=3/2$ state is a quartet $ \ket{JM_J}=\ket{3/2,-3/2}, \ket{3/2, -1/2}, \ket{3/2,1/2}, \ket{3/2,3/2}$. 

\section{Interactions}
\subsection{Lambda-neutron and lambda-proton interactions: GLM-YN0 potentials}

The $\Lambda p$ and $\Lambda n$ potentials used in this paper, called GLM-YN0 potentials, were
developed through Gel'fand-Levitan-Marchenko (GLM) inverse scattering theory~\cite{meo2019}
as simulations of the one-boson exchange NSC97f potential. The GLM-YN0 potentials are
charge-asymmetric and spin-dependent, meaning that $V^{(2S+1)}_{\Lambda p} \neq
V^{(2S+1)}_{\Lambda n}$, which reflects physical charge symmetry breaking in the $\Lambda N$
interaction. The $^1S_0$ and $^3S_1$ $\Lambda N$ potentials are fitted with three-range
Gaussians:
\begin{equation}
V^{(2S+1)}_{\Lambda p} = \sum_{i=1}^{3} U^{(2S+1)}_i
    \exp\!\left[-\frac{\left(r - \mu^{(2S+1)}_i\right)^2}
                      {\left(\sigma^{(2S+1)}_i\right)^2}\right]
\label{eq:VLp}
\end{equation}
\begin{equation}
V^{(2S+1)}_{\Lambda n} = \sum_{i=1}^{3} V^{(2S+1)}_i
    \exp\!\left[-\frac{\left(r - \nu^{(2S+1)}_i\right)^2}
                      {\left(\tau^{(2S+1)}_i\right)^2}\right]
\label{eq:VLn}
\end{equation}
The estimated parameters of these potentials are given in Tables~\ref{tab:fitlp}
and~\ref{tab:fitln}, respectively. We note that some shift parameters $\mu_i$ and $\nu_i$
take negative values (see for example $\mu^1_2 = -0.0795$~fm in Table~\ref{tab:fitlp}).
This is a feature of the GLM fitting procedure: displacing the Gaussian centres off the
origin provides the additional flexibility needed to accurately reproduce the short-range
repulsion and medium-range attraction of the NSC97f phase shifts with only three terms.

\begin{table}[h!]
\centering
\caption[]{Estimated parameters of spin-dependent $\Lambda$-proton potentials, $V_{\Lambda p}^{(2S+1)}$. For the potential $V_{\Lambda p}^{1}$ the $\chi^2/NDF=0.9976 $ and the p-value is 0.4948. For $V_{\Lambda p}^{3}$, the goodness-of-fit parameters are $\chi^2/NDF=0.9937 $ and p-value of 0.5094.}
\begin{tabular}{p{1.2cm}p{3.8cm}p{3.8cm}p{3.8cm}}
           &                   &   $V_{\Lambda p}^{1}$    &                               \\  \hline
           &  $U_{i}^{1}$ /MeV   &  $\mu_{i}^{1}$/ fm     &  $\sigma_{i}^{1}$/fm    \\ \hline \hline
$i=1$      & $135.4  $ &   $0.2615 \pm 0.001644$ &  $0.2765 \pm 0.001158$    \\ 
$i=2$      & $1774  $  &  $-0.07952 \pm 0.0006656 $ &  $0.1518 \pm 0.0003689 $\\ 
$i=3$      & $-144.6$  &   $0.389 \pm 0.0002739 $     &  $0.1531 \pm 0.0003456$       \\   \hline
           &                  &                               &                               \\
           &                  &                           &                               \\
           &                  &    $V_{\Lambda p}^{3}$                 &                               \\ \hline
           &  $U_{i}^{3}$  /MeV &   $\mu_{i}^{3}$/ fm                  &  $\sigma_{i}^{3}$/fm                \\ \hline \hline 
$i=1$      & $793.3  $   &  $ -0.3061 \pm 0.01053$ &  $ 0.2664 \pm 0.003863$       \\ 
$i=2$      & $-15.27  $ &  $ 0.3335 \pm 0.00113 $ &  $ 0.1722 \pm 0.00126  $        \\ 
$i=3$      & $2.421 $ &  $ 0.7139 \pm  0.002295$ &  $ 0.2044 \pm 0.002522 $       \\   \hline
\end{tabular}
\label{tab:fitlp}
\end{table}

\begin{table}[h!]
\centering
\caption[]{Estimated parameters of spin-dependent $\Lambda$-neutron potentials, $V_{\Lambda n}^{(2S+1)}$. For the potential $V_{\Lambda n}^{1}$ and $V_{\Lambda n}^{3}$, the $\chi^2/NDF$ and p-value are (0.9556, 0.6529) and (1.0406, 0.3388), respectively}
\begin{tabular}{p{1.2cm}p{3.8cm}p{3.8cm}p{3.8cm}}
           &                   &   $V_{\Lambda n}^{1}$       &                               \\  \hline
           &  $V_{i}^{1}$ /MeV &   $\nu_{i}^{1}$/ fm         &  $\tau_{i}^{1}$/fm                \\ \hline \hline
$i=1$      & $339.1  $ &   $ -0.1792 \pm 0.001442 $ &  $ 0.4561  \pm 0.00093 $      \\ 
$i=2$      & $6422  $  &   $ -0.12 \pm 0.001598 $   &  $ 0.1298 \pm 0.0001569 $       \\ 
$i=3$      & $-111.8 $ &   $ 0.3412 \pm 0.0004585  $  &  $ 0.1792 \pm 0.000612 $       \\   \hline
           &                   &                             &                               \\
           &                   &                             &                               \\
           &                   &    $V_{\Lambda n}^{3}$      &                               \\ \hline
           &  $V_{i}^{3}$ /MeV &   $\nu_{i}^{3}$/ fm         &  $\tau_{i}^{3}$/fm                \\ \hline \hline 
$i=1$      & $195.4  $   &  $ -0.1 \pm 0.0007632 $    &  $0.1865  \pm 0.0005573 $       \\ 
$i=2$      & $ -12.35  $ &  $ 0.3563 \pm  0.0007221 $ &  $0.1565 \pm 0.001315 $        \\ 
$i=3$      & $2.19  $    &  $ 0.7442 \pm 0.003085  $  &  $0.1872  \pm 0.004404 $       \\   \hline
\end{tabular}
\label{tab:fitln}
\end{table}

An important aspect of the $\Lambda N$ interaction is its spin dependence. Since the
GLM-YN0 potentials are $S$-wave ($L = 0$) potentials, there is no spin-orbit coupling.
Furthermore, since no higher partial waves are included, there is no angular momentum mixing
and the tensor force vanishes. The general $\Lambda p$ interaction therefore reduces to
\begin{equation}
V(r) = V_C(r) + V_\sigma(r)\,\vec{\sigma}_\Lambda \cdot \vec{\sigma}_p
\label{eq:Vgeneral}
\end{equation}
where $V_C(r)$ is the spin-independent central potential and $V_\sigma(r)$ is the spin-spin
potential. To determine $\vec{\sigma}_\Lambda \cdot \vec{\sigma}_p$ in each spin channel, we
use the total spin operator $\vec{S} = \vec{S}_\Lambda + \vec{S}_p$, where $\vec{S}_\Lambda =
\frac{1}{2}\vec{\sigma}_\Lambda$ and $\vec{S}_p = \frac{1}{2}\vec{\sigma}_p$. Expanding the
square of the total spin,
\begin{equation}
\vec{S}^{\,2} = \vec{S}^{\,2}_\Lambda + \vec{S}^{\,2}_p + 2\vec{S}_\Lambda \cdot \vec{S}_p
\end{equation}
and using $\vec{S}^{\,2}_\Lambda = \vec{S}^{\,2}_p = \tfrac{3}{4}$ together with
$\vec{S}_\Lambda \cdot \vec{S}_p = \tfrac{1}{4}\vec{\sigma}_\Lambda \cdot \vec{\sigma}_p$,
one obtains
\begin{equation}
\vec{\sigma}_\Lambda \cdot \vec{\sigma}_p = 2S(S+1) - 3
\label{eq:sigmadot}
\end{equation}
This gives $\vec{\sigma}_\Lambda \cdot \vec{\sigma}_p = -3$ in the singlet state
($^1S_0$,~$S = 0$) and $\vec{\sigma}_\Lambda \cdot \vec{\sigma}_p = +1$ in the triplet state
($^3S_1$,~$S = 1$). Substituting into Eq.~\eqref{eq:Vgeneral}, the singlet and triplet
potentials are, respectively,
\begin{equation}
V^1_{\Lambda p}(r) = V_C(r) - 3V_\sigma(r)
\label{eq:Vsinglet}
\end{equation}
\begin{equation}
V^3_{\Lambda p}(r) = V_C(r) + V_\sigma(r)
\label{eq:Vtriplet}
\end{equation}
Solving Eqs.~\eqref{eq:Vsinglet} and~\eqref{eq:Vtriplet} simultaneously for $V_C$ and
$V_\sigma$ gives the central and spin-spin components in terms of the fitted potentials:
\begin{equation}
V_C(r) = \frac{3}{4}V^3_{\Lambda p}(r) + \frac{1}{4}V^1_{\Lambda p}(r)
\label{eq:VC}
\end{equation}
\begin{equation}
V_\sigma(r) = \frac{1}{4}V^3_{\Lambda p}(r) - \frac{1}{4}V^1_{\Lambda p}(r)
\label{eq:Vsigma}
\end{equation}
The full operator form of the $\Lambda p$ interaction, which is the expression passed
directly to FaCE, is therefore
\begin{equation}
V_{\Lambda p}(r) = V_C(r) + V_\sigma(r)\,\vec{\sigma}_\Lambda \cdot \vec{\sigma}_p
\label{eq:VLpfull}
\end{equation}
with $V_C$ and $V_\sigma$ as given in Eqs.~\eqref{eq:VC} and~\eqref{eq:Vsigma}. It can be verified that this recovers the original fitted potentials: in the singlet channel
$\vec{\sigma}_\Lambda \cdot \vec{\sigma}_p = -3$, so the spin-spin contribution is
$-3V_\sigma(r)$ and Eq.~\eqref{eq:VLpfull} reduces to $V^1_{\Lambda p}$; in the triplet
channel $\vec{\sigma}_\Lambda \cdot \vec{\sigma}_p = +1$, so the contribution is $+V_\sigma(r)$
and Eq.~\eqref{eq:VLpfull} reduces to $V^3_{\Lambda p}$, as required. Following an identical
procedure, the $\Lambda n$ interaction is
\begin{equation}
V_{\Lambda n}(r) =
    \left[\frac{3}{4}V^3_{\Lambda n}(r) + \frac{1}{4}V^1_{\Lambda n}(r)\right]
  + \left[\frac{1}{4}V^3_{\Lambda n}(r) - \frac{1}{4}V^1_{\Lambda n}(r)\right]
    \vec{\sigma}_\Lambda \cdot \vec{\sigma}_n
\label{eq:VLnfull}
\end{equation}
where $V^{1,3}_{\Lambda n}$ are taken from Table~\ref{tab:fitln}. Since the GLM-YN0 potentials
are charge-asymmetric, $V^{(2S+1)}_{\Lambda n} \neq V^{(2S+1)}_{\Lambda p}$, so both
interactions must be treated independently in the three-body calculation. In the original NSC97f paper, the lambda-nucleon potentials have charge-symmetry breaking that arises from $\Lambda - \Sigma$ mixing. In the application of inverse scattering theory in \cite{meo2019}, the NSC97f scattering phases used as input do have CSB encoded in them. This CSB effect manifests in the simulated potentials in \cite{meo2019}: the lambda-proton potentials are more attractive than the lambda-neutron potentials and in both cases the spin singlet potentials are more attractive than the spin triplet potentials.

\subsection{Nucleon-nucleon interactions}

Since both the neutron and proton each have spin 1/2 and isospin 1/2, angular momentum algebra allows the following spin-isospin channels in a neutron-proton interaction: $(S_{np}, I_{np}) =(0,0), (0,1), (1,0), (1,1)$. For S-waves, antisymmetrisation requirements on the total wavefunction restrict the system only to the (0,1), (1,0) channels i.e. the spin singlet channel (appears in np scattering data) and the spin-triplet channel (appears in scattering data and in the deuteron). A number of hypertriton computations leave out the spin singlet contribution to the np interaction, and in some cases the hypertriton is treated as a $\Lambda$-deuteron two-body problem \cite{con1992, ber2023}. Even though the spin-singlet channel contributes very little to the binding, it is important to include it for high precision. Furthermore, if other observables besides binding energy are to be computed e.g. radii, it is more accurate to include the singlet channel because some observables may need to probe this channel of the interaction. In this paper we explicitly include contributions from both channels. We employ the (0,1) and (1,0) semi-realistic Malfliet-Tjon I/III np potentials from~\cite{fri1990}. Following the same methodology as in the previous subsection, the central and spin-spin parts of the np potential are given by

\begin{equation}
V_{n p}(r) =
    \left[\frac{3}{4}V^3_{np}(r) + \frac{1}{4}V^1_{np}(r)\right]
  + \left[\frac{1}{4}V^3_{np}(r) - \frac{1}{4}V^1_{np}(r)\right]
    \vec{\sigma}_n \cdot \vec{\sigma}_p
\label{eq:Vnp}
\end{equation}

\section{Results and discussion}
\subsection{$\Lambda np(1/2^+)$}

The ground state of the hypertriton ($J^\pi = 1/2^+$) was investigated using the hyperspherical harmonic expansion of the Faddeev equations and the GLM-YN0 $\Lambda N$ potentials together with the Malfliet-Tjon I/III $np$ interaction. The three-body wavefunction is built from three Faddeev components, each expressed in its own Jacobi (spectator) coordinate set: X (spectator $\Lambda$, interacting pair $np$), Y (spectator $n$, interacting pair $\Lambda p$), and T (spectator $p$, interacting pair $\Lambda n$). Within each component the wavefunction is expanded on a hyperspherical channel basis. 

In the FaCE code \cite{tho2004}, the model space is delimited by the size of the Laguerre basis ($N_{bmax}$ for the hyperradial part, the number of Jacobi polynomials $N_{jac} $ for the hyperangular part and the quantum numbers $K_{max} $, $l_x$ and $l_y$. The number of Jacobi polynomials used is $N_{jac}=180$, together with the quantum number cutoffs $l_{xmax} =2.0$ and $l_{ymax} =2.0$ and $s_{xmax}=1.0 $. Convergence of the three-body binding energy was assessed with respect to two independent parameters: the hyperangular momentum cutoffs for the np subsystem $K_{np}^{max}$, the lambda-nucleon subsystems $K_{\Lambda N}^{max}$, and $N_b^{max}$. Since the present implementation of FaCE cannot extend beyond $K_{\Lambda N}^{max}=12$ for the $\Lambda N$ subsystem channels because of memory related problems, we adopt a hierarchical scheme: the $np$ sector converges rapidly and is fixed at $K_{np}^{max}=18$, well beyond the point of sensitivity, while $K_{\Lambda N}^{max}$ is scanned from 2 to 12. Segmentation faults or memory related issues \cite{kha2012} and slow convergence \cite{bac2009, bac2012, sch1972} are well-known issues in hyperspherical harmonics calculations .

Table~\ref{tab:nbmax3} shows the resulting ground-state energy at three Laguerre basis sizes, $N_b^{max}=30$, $50$, and $70$, computed to confirm that the basis itself had saturated.

\begin{table}[htbp]
\centering
\caption{Ground-state energy of $\Lambda np(1/2^+)$ as a function of
$K_{\Lambda N}^{max}$ for three Laguerre basis sizes, with $K_{np}^{max}=18$
fixed throughout.}
\label{tab:nbmax3}
\begin{tabular}{cccc}
\hline
$K_{\Lambda N}^{max}$ & $E$ ($N_b^{max}=30$, MeV) & $E$ ($N_b^{max}=50$, MeV) & $E$ ($N_b^{max}=70$, MeV) \\
\hline
2  & $-2.334087$ & $-2.334677$ & $-2.334682$ \\
4  & $-2.334697$ & $-2.335290$ & $-2.335294$ \\
6  & $-2.333940$ & $-2.334526$ & $-2.334530$ \\
8  & $-2.334280$ & $-2.334868$ & $-2.334872$ \\
10 & $-2.334393$ & $-2.334982$ & $-2.334985$ \\
12 & $-2.334381$ & $-2.334969$ & $-2.334973$ \\
\hline
\end{tabular}
\end{table}

A pseudostate, which is an artefact of the basis, would be expected to show a strong sensitivity to $N_b^{max}$. The small shift observed here is the signature of a genuine bound state. We fit the $N_b^{max}=70$ series to an exponential-decay form, $E(K_{\Lambda N}^{max}) = E_\infty + A\,e^{-bK_{\Lambda N}^{max}}$, the convergence pattern more commonly reported for hyperspherical-harmonics calculations (see Ref.~\cite{bac2009}). The fit yields $E_\infty = -2.335 \pm 0.0003$ MeV. We therefore take $K_{\Lambda N}^{max}=12$ as the point of practical convergence and adopt $E=-2.335 \pm 0.0003$ MeV as our final ground-state energy of the hypertriton. 

From the deuteron threshold of the same Malfliet-Tjon I/III potential ($E(^2\mathrm{H}) = -2.2307$ MeV), this corresponds to a $\Lambda$ separation energy of

\begin{equation}
B_\Lambda(^3_\Lambda\mathrm{H}) = E(^3_\Lambda\mathrm{H}) - E(^2\mathrm{H}) = 2.335 - 2.2307  \approx 0.104~\mathrm{MeV}.
\label{eq:blambda_new}
\end{equation}

Without stating any specific value, Ref. \cite{miy1999} affirms that the original NSC97f potentials reproduce the correct experimental binding at that time i.e. $B_{\Lambda} =0.13 \pm 0.05$ MeV. In Ref. \cite{hiy2014}, using Gaussian functions that accurately simulate the original NSC97f phase shifts, a lambda binding energy of $B_{\Lambda}=0.19 $ MeV was reported within the Gaussian Expansion Method. Using the same NSC97f-simulated potential as Ref. \cite{hiy2014}, a value of $B_{\Lambda}=0.17 $ MeV was reported in Ref. \cite{fer2017} for the Non-Symmetrized Hyperspherical Harmonics method. 

The wave function decomposition is summarized in Table~\ref{tab:faddeev_norms}, while the partial-wave decomposition is shown in Table~\ref{tab:partial_wave}. Since the three Jacobi trees are related by a geometric (Raynal--Revai) rotation rather than a simple coordinate relabelling, they are not mutually orthogonal, so these three component norms do not need to sum to 1.

\begin{table}[ht]
\centering
\caption{Faddeev component norms for the $\Lambda np(1/2^+)$ ground state.}
\label{tab:faddeev_norms}
\begin{tabular}{lcc}
\hline
Component & Norm & Notes \\
\hline
X (dominant) & 1.000000 & $\Lambda$--(np) configuration, spectator $\Lambda$ \\
Y & 0.381677 & Spectator neutron \\
T & 0.381978 & Spectator proton \\
\hline
\end{tabular}
\end{table}

\begin{table}[h]
\centering
\caption{Partial-wave decomposition of the $\Lambda np(1/2^+)$ ground state.}
\label{tab:partial_wave}
\begin{tabular}{lcccl}
\hline
Component & $X$ (Probability) & $Y$ (Probability) & $T$ (Probability) & Notes \\
\hline
$P(S)$  & 0.999995 & 0.079818 & 0.079939 & Symmetric $S$-state \\
$P(S')$ & 0.000004 & 0.301871 & 0.302051 & Mixed-symmetry $S$-state \\
$P(P)$  & 0.000000 & 0.000000 & 0.000000 & $P$-wave \\
$P(D)$  & 0.000000 & 0.000000 & 0.000000 & $D$-wave \\
\hline
\end{tabular}
\end{table}

The X component, with a norm of essentially 1.000000, dominates the three-body wave function. Its wavefunction is overwhelmingly of S-state character: $\mathrm{P}(S) = 0.999995$, $\mathrm{P}(S') = 0.000004$, $\mathrm{P}(P) \approx 0$, $\mathrm{P}(D) \approx 0$. Higher partial waves (P- and D-waves) are virtually absent. The pure S state dominates this channel, reflecting the dominance of the $^3S_1$ $np$ i.e. the deuteron. The small mixed-S probability reflects the weak admixture of the $^1S_0$ neutron–proton configuration allowed in the $J^{\pi}  = 1/2^+$ state through spin coupling. 

In contrast, the Y and T components contain a considerably larger mixed-symmetry S-state probability than the X component. Since these components describe the $\Lambda N$ subsystems, they reflect the coupling between the singlet and triplet $\Lambda N$ spin channels induced by the full three-body dynamics. The charge-symmetry breaking between the $\Lambda p$ and $\Lambda n$ interactions produces a small numerical difference between the Y and T probabilities.

The Y and T components' features may be understood as arising from the stronger spin dependence of the $\Lambda N$ interaction driven by a large difference in short-range repulsion, and the difference in attraction strength between the $\Lambda p$ and $\Lambda n$ interactions. In the X-component, the np triplet state is favoured because it supports a bound state i.e. it minimises the energy of the hypertriton. For the lambda-nucleon subsystems, neither the triplet nor the singlet supports a bound state in current lambda-nucleon interaction models. 

\subsection{$\Lambda np(3/2^+)$}

The $J^\pi = 3/2^+$ state of the $\Lambda np$ system was investigated using the same hyperspherical harmonic expansion of the Faddeev equations and the GLM-YN0 $\Lambda N$ potentials together with the Malfliet--Tjon I/III $np$ interaction, following the same three-Faddeev-component decomposition used in the $J^\pi = 1/2^+$ state. 

The same hierarchical convergence scheme was applied to the $J^\pi=3/2^+$ state, with $K_{np}^{max}=18$ fixed and $K_{\Lambda N}^{max}$ varied for three Laguerre basis sizes,
$N_b^{max}=30$, $50$, and $70$. Segmentation faults prevented calculations beyond $K_{\Lambda N}^{max}=6$ for this channel, and the $K_{\Lambda N}^{max}=0$ point is omitted. This leaves the series $K_{\Lambda N}^{max}=2,4,6$.

\begin{table}[htbp]
\centering
\caption{Energy of the $\Lambda np(3/2^+)$ state as a function of
$K_{\Lambda N}^{max}$ for three Laguerre basis sizes, with
$K_{np}^{max}=18$ fixed. Calculations beyond $K_{\Lambda N}^{max}=6$
were not possible due to segmentation faults.}
\label{tab:nbmax32}
\begin{tabular}{cccc}
\hline
$K_{\Lambda N}^{max}$ & $E$ ($N_b^{max}=30$, MeV) & $E$ ($N_b^{max}=50$, MeV) & $E$ ($N_b^{max}=70$, MeV) \\
\hline
2 & $-0.784730$ & $-0.785542$ & $-0.785544$ \\
4 & $-0.785949$ & $-0.786760$ & $-0.786763$ \\
6 & $-0.785942$ & $-0.786752$ & $-0.786755$ \\
\hline
\end{tabular}
\end{table}

With only three usable points we apply two independent extrapolation methods: the exponential-decay Ansatz $E(K_{\Lambda N}^{max}) = E_\infty + A\,e^{-bK_{\Lambda N}^{max}}$ of
Ref.~\cite{bac2009}, and the Aitken $\Delta^2$ process,

\begin{equation}
E_\infty \approx E_n - \frac{(E_{n+1}-E_n)^2}{E_{n+2}-2E_{n+1}+E_n},
\end{equation}

applied to the only available set of three points $(K_{\Lambda N}^{max}=2,4,6)$ at each basis size. Both methods use all available degrees of freedom exactly (three points, three parameters
for the exponential fit; a single triple for Aitken), so neither yields a residual-based uncertainty; agreement between the two independent methods is used instead as a consistency check.

\begin{table}[htbp]
\centering
\caption{Extrapolated $\Lambda np(3/2^+)$ energy from the exponential-decay
fit and the Aitken $\Delta^2$ process, both applied to the
$K_{\Lambda N}^{max}=2,4,6$ points.}
\label{tab:expfit32}
\begin{tabular}{ccc}
\hline
$N_b^{max}$ & $E_\infty$, exponential fit (MeV) & $E_\infty$, Aitken $\Delta^2$ (MeV) \\
\hline
30 & $-0.7859$ & $-0.7859$ \\
50 & $-0.7868$ & $-0.7868$ \\
70 & $-0.7868$ & $-0.7868$ \\
\hline
\end{tabular}
\end{table}

The two extrapolation methods agree at every basis size, which is reassuring given that neither can otherwise be cross-checked against a residual. The extrapolated energy shifts by
$\sim0.81$ keV from $N_b^{max}=30$ to $50$, but by only $\sim0.003$ keV from $N_b^{max}=50$ to $70$. This is the same pattern of basis saturation observed for the $J^\pi=1/2^+$ ground state, where the $50\to70$ shift was likewise negligible compared to the $30\to50$ shift. This consistency across extrapolation methods and across the two $J^\pi$ channels, despite
the more limited $K_{\Lambda N}^{max}$ range available here, supports $E=-0.787$ MeV as a reliable estimate for the $J^\pi=3/2^+$ state, though we note this result carries a larger uncertainty than the ground-state value given the restricted $K_{\Lambda N}^{max}$ range
imposed by the segmentation fault.

Even though this state is bound with respect to the $\Lambda + n + p$ threshold, it is unbound relative to the $\Lambda + d$ threshold, from the deuteron binding energy of the same Malfliet--Tjon I/III potential, $E_d = -2.2307$ MeV. The $\Lambda$ separation energy is

\begin{equation}
B_\Lambda = E - E_d = 0.787 - 2.2307  \approx -1.444\ \text{MeV},
\end{equation}

This indicates that the $J^\pi=3/2^+$ state is unbound relative to the $\Lambda+d$ breakup threshold by approximately 1.444 MeV under the present GLM-YN0 interaction model. In Ref. \cite{sch2020}, it was found that $\Lambda np(3/2^+)$ is a near-threshold virtual state. In a follow-up paper \cite{sch2022}, it was observed that increasing the strength of the lambda-nucleon force caused this virtual state to move closer to the $\Lambda d$ threshold, becoming a shallow bound state. The FaCE methodology can only compute bound states or pseudostates. The most robust conclusion is that the $J^\pi=3/2^+$ state is unbound. Whether it is a virtual state or a resonance, cannot be handled in this code. The results are summarised in Table \ref{tab:hypertriton_thresholds}. In line with Refs. \cite{oga1980, miy1995} we find the $\Lambda np(3/2^+)$ state unbound. Refs. \cite{sch2020, sch2022} go further to classify this unbound state as a virtual state. 

\begin{table}[h]
\centering
\begin{tabular}{c|c}
\hline
Channel & $E$ (MeV) \\
\hline
$\Lambda+n+p$                        & 0       \\
${}^{3}_{\Lambda}\mathrm{H}(3/2^+)$  & $-0.787$  \\
$\Lambda+d$                          & $-2.2307$ \\
${}^{3}_{\Lambda}\mathrm{H}(1/2^+)$  & $-2.335$  \\
\hline
\end{tabular}
\caption{Energy levels of the hypertriton system relative to the
three-body $\Lambda+n+p$ threshold. The $J^\pi=3/2^+$ state lies above
the $\Lambda+d$ breakup threshold and therefore is not a bound excited
hypertriton state.}
\label{tab:hypertriton_thresholds}
\end{table}

\section{Conclusions}

We have computed the $\Lambda np$ three-body system in the $J^\pi = 1/2^+$ and $3/2^+$ channels using a hyperspherical-harmonic expansion of the Faddeev equations. These calculations were carried out with $\Lambda p$ and $\Lambda n$ potentials constructed via Gel'fand--Levitan--Marchenko inverse scattering theory. For the ground state we find a binding energy of $-2.335$~MeV, corresponding to $B_\Lambda = 0.104$~MeV, appreciably below the recent MAMI/A1 value of $0.523 \pm 0.013_{\text{stat}} \pm 0.075_{\text{syst}}$~MeV and the STAR value, though closer in magnitude to the shallower J-PARC E07 and ALICE results. The $J^\pi = 3/2^+$ state lies 1.444~MeV above the $\Lambda + d$ breakup threshold and is therefore unbound, as reported in other studies.

These results confirm that inverse scattering theory provides a genuine and workable alternative to $G$-matrix methods for simulating $\Lambda N$ interactions. By construction, the GLM-YN0 potentials are exactly phase-equivalent to the input NSC97f phase shifts, thereby capturing all NSC97f on-shell content exactly, and some off-shell behaviour. This distinction matters most in the hyperon sector, where scattering data is too sparse to tightly constrain a potential in the first place, making phase-equivalence to the available scattering data an added advantage. A natural continuation of this work is to extend the inverse scattering method beyond the S-wave channels by performing a coupled-channel Gel'fand–Levitan–Marchenko inversion that incorporates higher partial waves, including the P- and D-wave interactions.


\end{document}